\documentclass[a4paper]{spie}  

\usepackage{amsmath,amsfonts,amssymb}
\usepackage{graphicx}
\usepackage{multirow}
\usepackage[colorlinks=true, allcolors=blue]{hyperref}
\usepackage[top=2.54cm,bottom=4.94cm,left=1.925cm,right=1.925cm]{geometry}

\title{Prime Focus Spectrograph (PFS) for the Subaru Telescope:
  Overview, recent progress, and future perspectives}

\author[a]{Naoyuki Tamura}
\affil[a]{Kavli Institute for the Physics and Mathematics of the Universe (WPI),The University of Tokyo Institutes for Advanced Study, The University of Tokyo, Kashiwa, Chiba 277-8583, Japan}
\author[b]{Naruhisa Takato}
\affil[b]{Subaru Telescope, National Astronomical Observatory of Japan, 650 North A'ohoku Place, Hilo, HI 96720, USA}
\author[a]{Atsushi Shimono}
\author[a]{Yuki Moritani}
\author[a]{Kiyoto Yabe}
\author[a]{Yuki Ishizuka}
\author[c]{Akitoshi Ueda}
\author[c]{Yukiko Kamata}
\affil[c]{National Astronomical Observatory of Japan, 2-21-1 Osawa, Mitaka, Tokyo 181-8588, Japan}
\author[d]{Hrand Aghazarian}
\affil[d]{Jet Propulsion Laboratory, 4800 Oak Grove Dr., Pasadena, CA 91109, USA}
\author[e]{St\'{e}phane Arnouts}
\affil[e]{Aix Marseille Universit\'{e}, CNRS, LAM (Laboratoire d'Astrophysique de Marseille) UMR 7326, 13388, Marseille, France}
% ...
\authorinfo{Further author information: (Send correspondence to Naoyuki Tamura.)\\E-mail: naoyuki.tamura@ipmu.jp, Telephone: $+$81 (0)4 7136 6531}

\begin{document} 
\maketitle

\begin{abstract}

PFS (Prime Focus Spectrograph), a next generation facility instrument
on the 8.2-meter Subaru Telescope, is a very wide-field, massively
multiplexed, optical and near-infrared spectrograph. Exploiting the
Subaru prime focus, 2394 reconfigurable fibers will be distributed
over the 1.3 deg field of view. The spectrograph has been designed
with 3 arms of blue, red, and near-infrared cameras to simultaneously
observe spectra from 380nm to 1260nm in one exposure at a resolution
of $\sim$1.6$-$2.7\AA. An international collaboration is developing
this instrument under the initiative of Kavli IPMU. The project is now
going into the construction phase aiming at undertaking system
integration in 2017-2018 and subsequently carrying out engineering
operations in 2018-2019. This article gives an overview of the
instrument, current project status and future paths forward.

\end{abstract}


\section{INTRODUCTION}
\label{sec:intro}  

The wide-field capability at the prime focus is clearly one of the key
advantages of the 8.2m Subaru Telescope, and a few instruments are
exploiting this to deliver valuable scientific data. Suprime-Cam
(Subaru Prime Focus Camera)\cite{miyazaki02} is a wide-field imager
with a mosaic of ten 2K$\times$4K CCDs covering a field of
$34^{\prime} \times 27^{\prime}$. Its broad-band and narrow-band deep
and wide imaging data have been powerful for a number of discoveries
and detailed characterization of various astronomical objects over a
wide range of redshift. FMOS (Fiber Multi-Object
Spectrograph)\cite{kimura10, tamura12} is a wide-field, fiber-fed
multi-object spectrograph, where the 400 fibers on the half-degree
field are reconfigurable with an Echidna-style fiber positioner
system\cite{jurek04}. The spectra cover the near-infrared regime from
900nm to 1800nm. This allows highly efficient observation of the
rest-frame optical spectral features for objects at redshifts beyond
one (e.g. \cite{yabe14}) and absorption bands in the infrared
continuum of cool, low-mass stars (e.g. \cite{muzic12}). Recently, the
Subaru Telescope observatory has started accepting a large program in
the framework called ``Subaru Strategic Program'' (SSP) that can span
up to $\sim$300 nights over $\sim$5 years. A cosmology survey program
called ``FastSound''\cite{tonegawa15} (PI: T. Totani), one of the few
SSP programs, was successfully completed with FMOS to reveal a 3D map
of $\sim$3000 galaxies around $z\sim1.4$ and a significant detection
of Redshift Space Distortions (RSD)\cite{okumura16}.

Now new instrumentation projects are underway to upgrade the Subaru
prime focus and push the cutting edge science further forward, taking
the full advantage of the unique wide field of view. Hyper Suprime-Cam
(HSC)\cite{miyazaki12}, the successor of Suprime-Cam, is a very
wide-field imager with a 1.5-degree diameter field of view ``paved''
by 116 2K$\times$4K CCDs. It has been in science operation since 2014
and a 5-year, 300-night SSP survey program is on-going. PFS (Prime
Focus Spectrograph), as described in this article, is a very
wide-field, massively multiplexed, optical and near-infrared (NIR)
spectrometer.  The focal plane will be equipped with 2394
reconfigurable fibers distributed in the 1.3-degree wide hexagonal
field of view. The spectrograph has been designed to cover a wide
range of wavelengths simultaneously from 380nm to 1260nm in one
exposure. The PFS and HSC instrumentation projects are under the
umbrella of the Subaru Measurement of Images and Redshifts (SuMIRe)
project (PI: H. Murayama) aiming to conduct deep and wide sky surveys
exploiting the unique capability of the Subaru Telescope. It should be
emphasized that HSC and PFS enable deep imaging and spectroscopic
surveys of the same region of sky using the same 8.2m telescope,
allowing one to have good understandings of various systematics in the
data.

\begin{figure}
  \begin{center}
       \includegraphics[width=8cm]{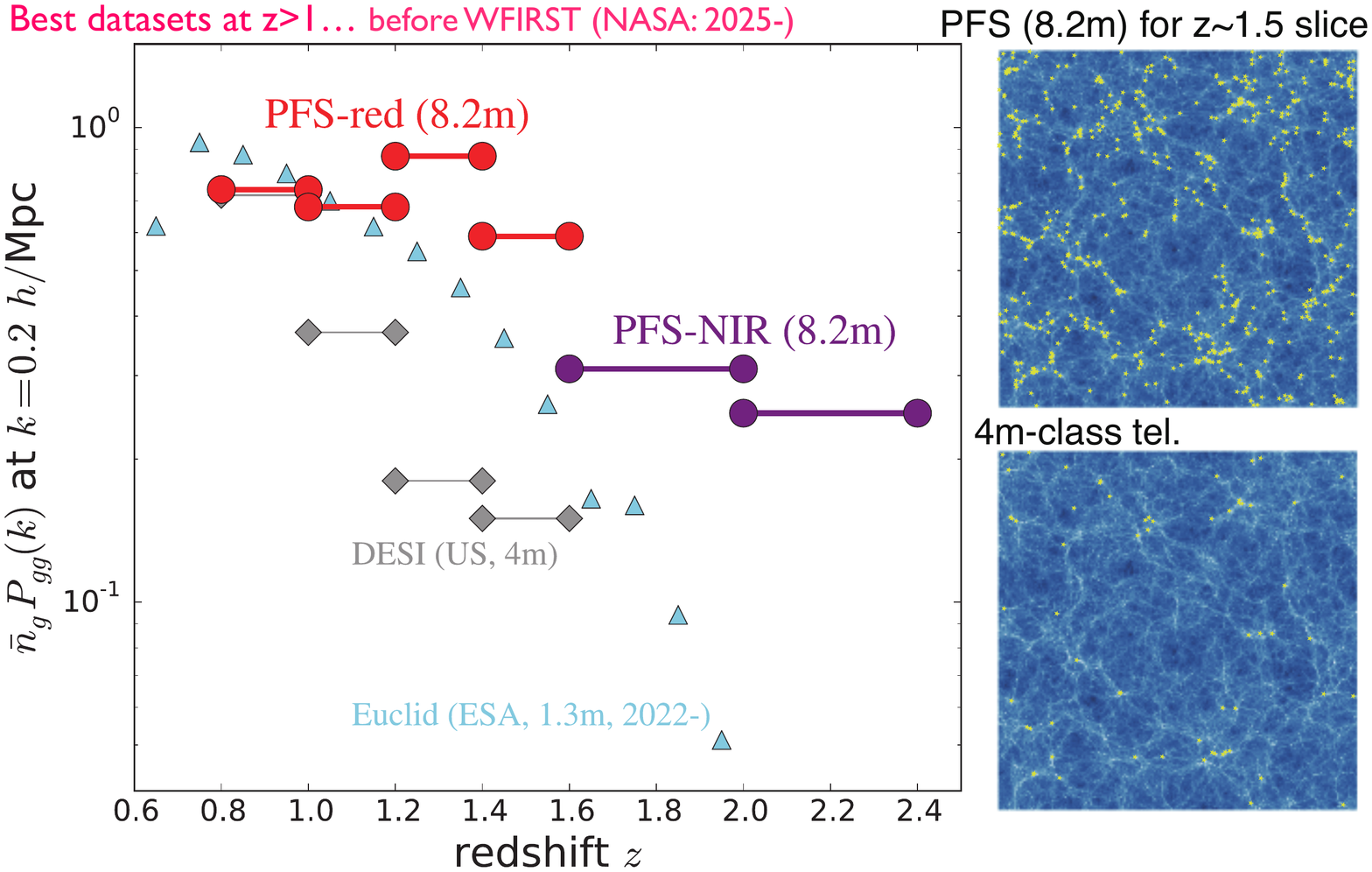}
       \includegraphics[width=8cm]{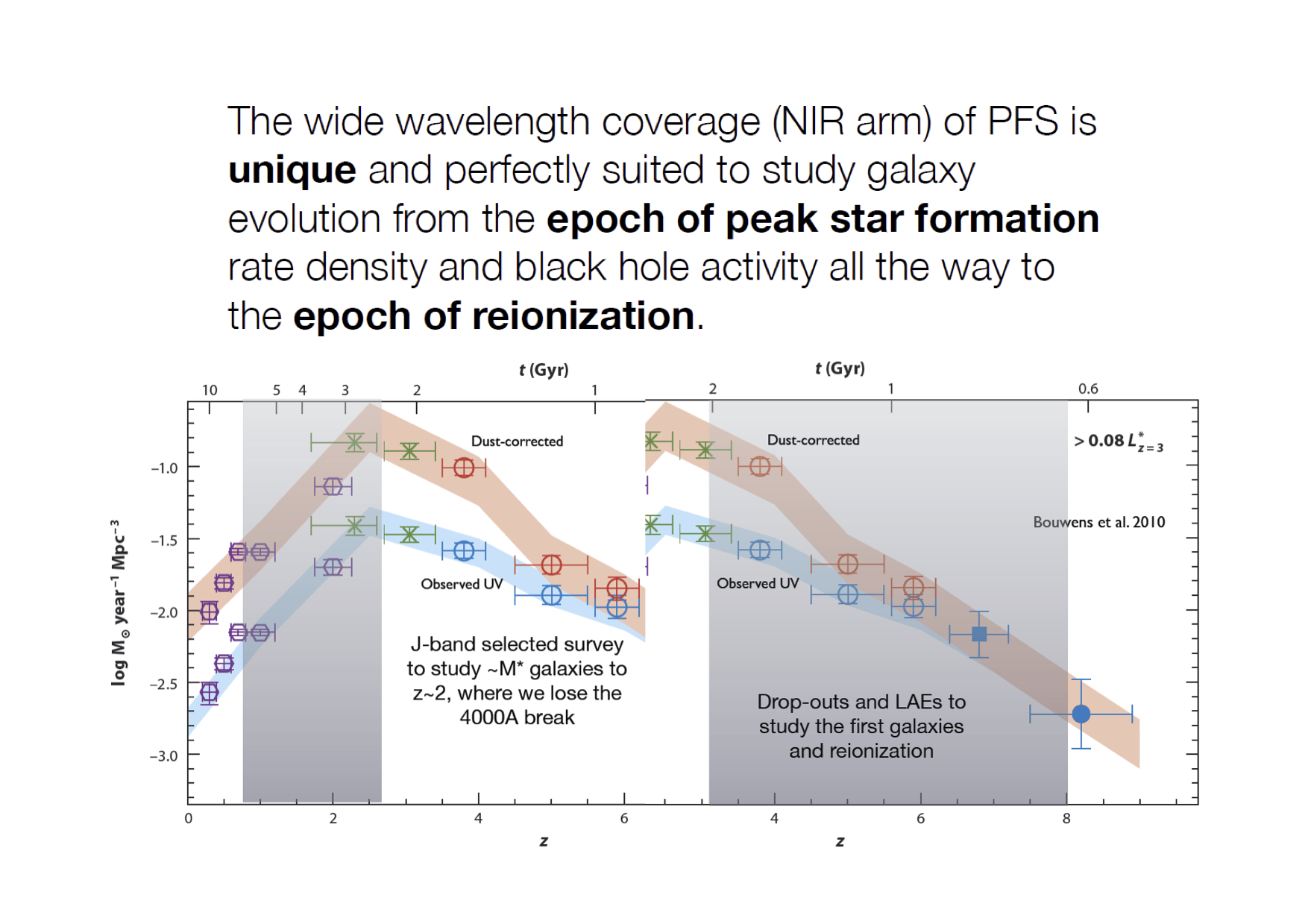}
       
       \includegraphics[width=8cm]{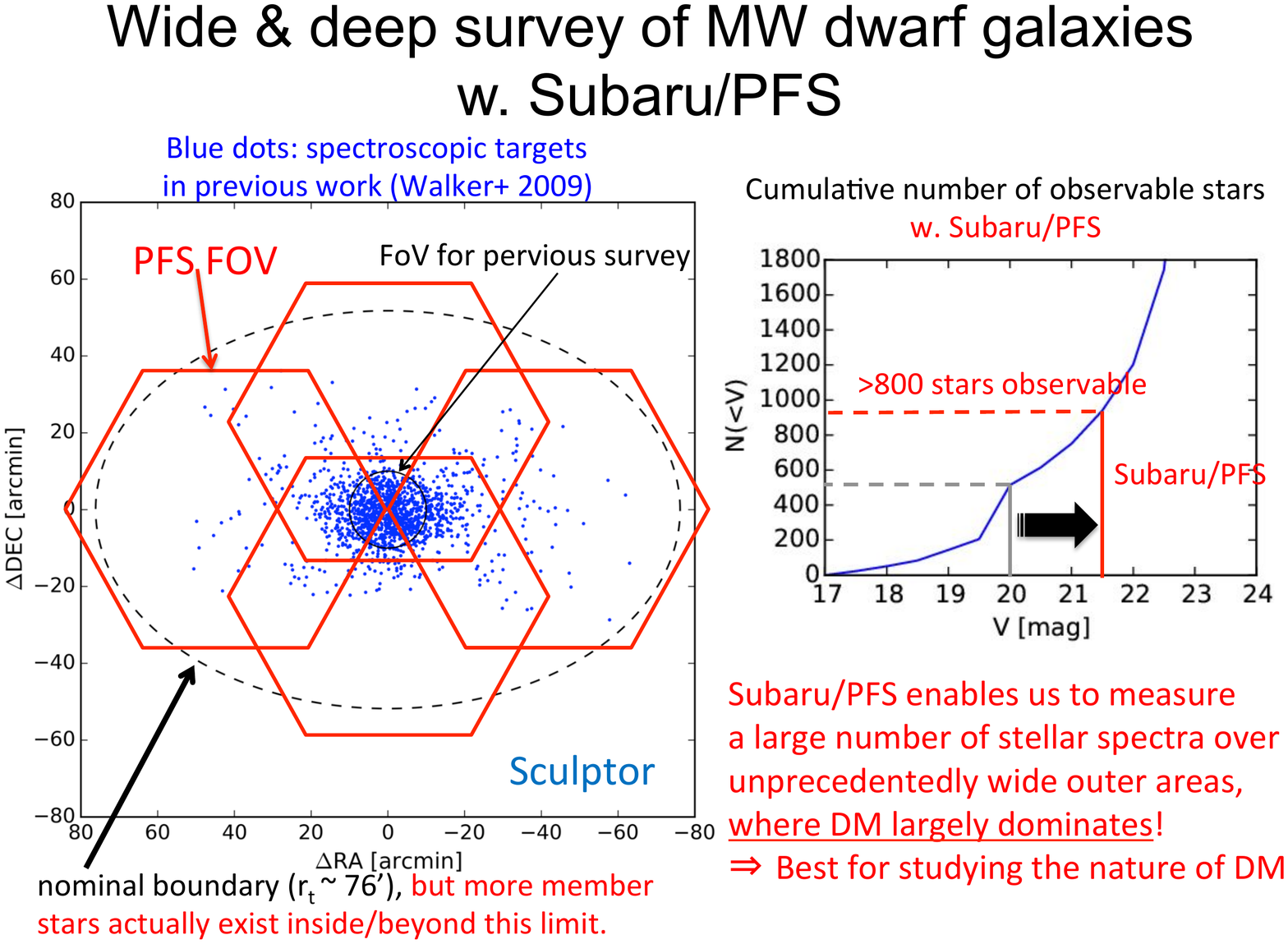}
  \end{center}
  \caption{ \label{fig:strength} A few representations of the PFS
    strengths. {\it Top left:} the product of the mean galaxy space
    density and the galaxy clustering amplitude is plotted as a
    function of redshift and the case for the PFS cosmology program is
    compared with WFIRST and DESI\cite{np-desi,np-wfirst-euclid}. {\it
      Top right:} the cosmic star formation history\cite{csfr} is
    illustrated with a wide range of redshifts that can be accessed in
    the PFS galaxy \& AGN evolution survey exploiting the wide
    wavelength coverage. {\it Bottom:} a PFS pointing plan around the
    Sculptor dwarf galaxy\cite{dsph1, dsph2} as a part of the Galactic
    archaeology program is summarized, showing high efficiency thanks
    to the wide field and high multiplicity.}
\end{figure}

Envisioning a large survey in the SSP framework, the PFS science team
has built a preliminary survey plan and has developed top-level
requirements for the instrument\cite{takada14}. The goal is to address
key questions in three main fields: cosmology, galaxy \& AGN
evolution, and Galactic archaeology, and from the joint implications,
to understand the dark sector of the universe. The team has been
continuously refining the plan as the instrument characteristics and
technical constraints on the survey observation process are better
understood. The combination of the wide field, high multiplicity, and
high number density of the fibers on the focal plane offers an
opportunity of designing a unique survey on these three core science
cases envisioned in the PFS SSP survey (Fig. \ref{fig:strength}).

\begin{figure}
  \begin{center}
    \includegraphics[width=15cm]{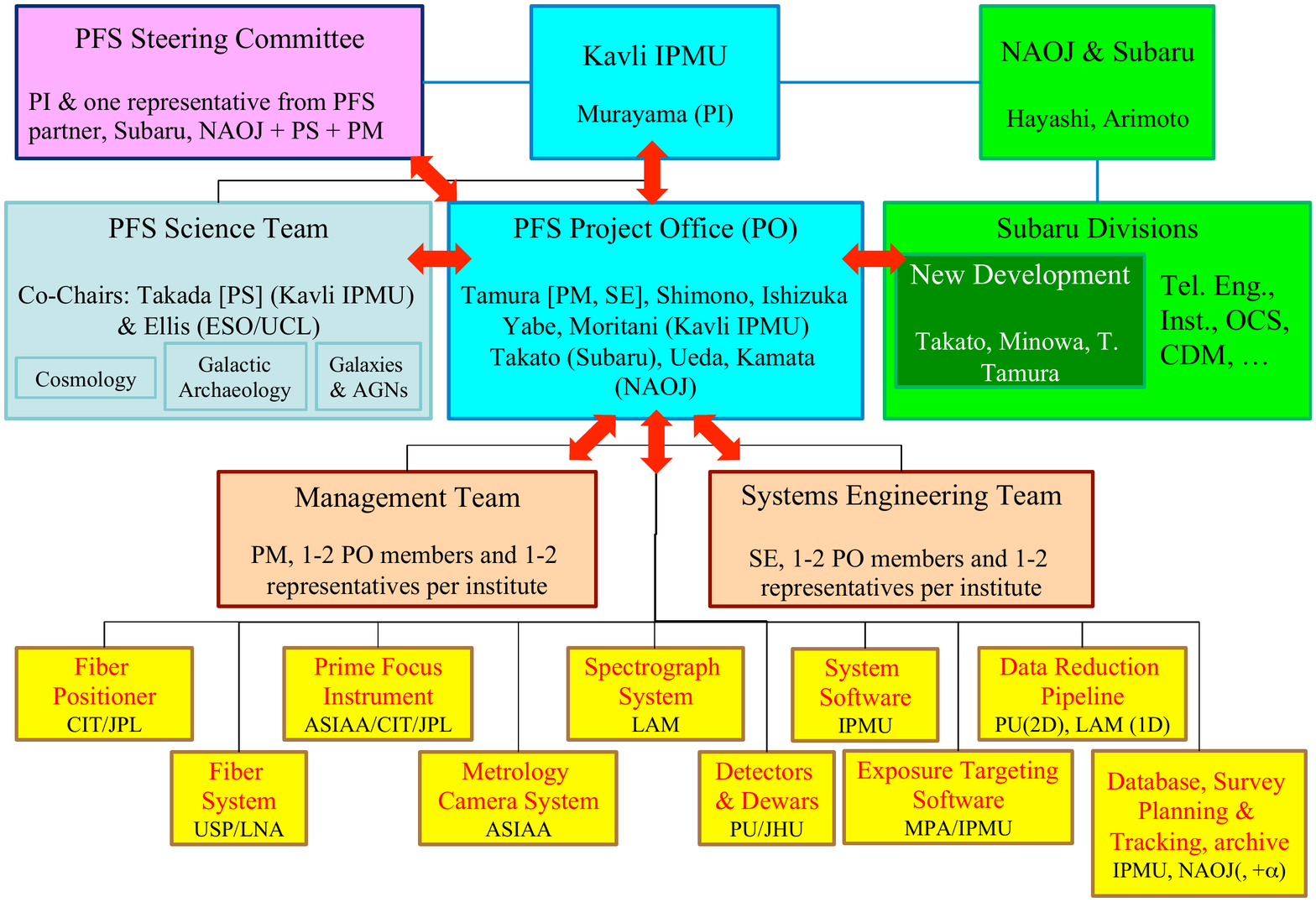}
  \end{center}
  \caption{ \label{fig:org} A chart showing the PFS organization and
    the assignments of instrument subsystems and subcomponents to the
    collaboration.}
\end{figure}

The development of this instrument has been undertaken by an
international collaboration at the initiative of Kavli IPMU, with work
packages for subsystem and subcomponent development assigned to
various collaborating institutions (Fig. \ref{fig:org}). The project
is now in the phase of construction, integration and test aiming to
start science operations from mid-late 2019. In parallel, detailed
modeling of the instrument and output spectral images are on-going in
order to characterize the instrument on-sky capabilities and
accordingly finalize the SSP survey design. This way, a PFS SSP
program for follow-up spectroscopy can start in a timely manner
subsequently after the HSC SSP survey. In what follows, the instrument
basics are described in \S~\ref{sec:inst}, and an overview of the
instrument and survey operation concept is given in
\S~\ref{sec:operation}. Then updates of a few major aspects of the
project are summarized in \S~\ref{sec:updates}, and this article is
summarized and a timeline for the future developments is given in
\S~\ref{sec:future}.

\section{THE INSTRUMENT}
\label{sec:inst}

\begin{figure}
  \begin{center}
    \includegraphics[width=18cm]{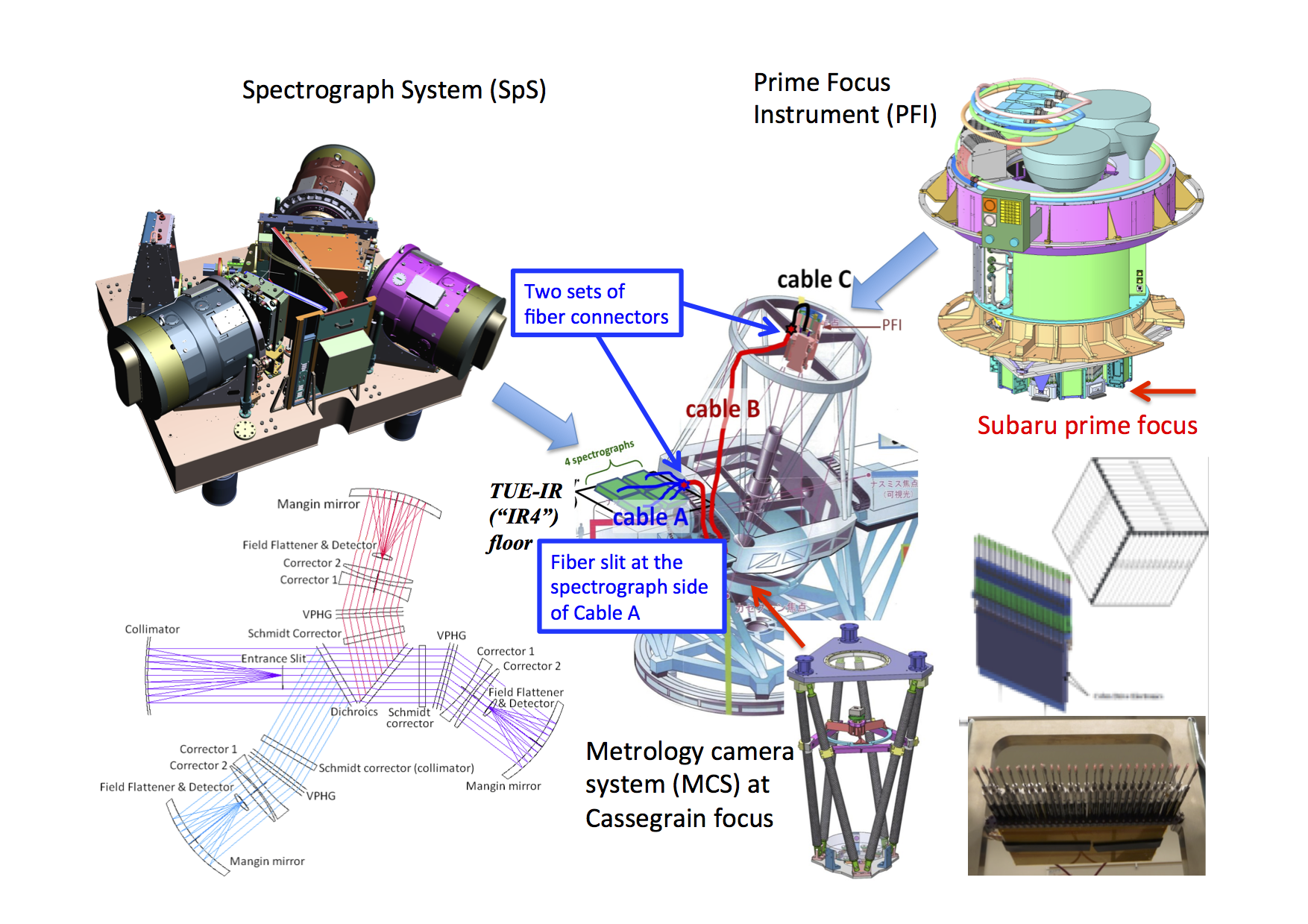}
  \end{center}
  \caption{ \label{fig:inst} A schematic view of the configuration of
    PFS subsystems. An overall sketch of the Subaru Telescope is
    presented in the middle with the PFS fiber cable routed from the
    prime focus to the spectrograph system. On the right, a solid
    model of PFI (top), a schematic view of the focal plane (middle),
    and a photo of the Cobra engineering model fiber positioners
    module are presented. On the left, a solid model of one
    spectrograph module (top) and a ray-trace view of it (bottom) are
    shown.}
\end{figure}

The PFS instrument is composed of four subsystems, whose distribution
on the telescope is illustrated in Fig. \ref{fig:inst}: The lights
from astronomical objects and sky are fed to the fibers configured at
the Subaru prime focus, are then transmitted via the fiber cable to
the spectrograph system in the telescope enclosure building, and the
spectral images of them are delivered on the spectrograph
detectors. We here give an overview of these subsystems. The
instrumental software system and survey operations are described in
\S~\ref{sec:operation}.

At the Subaru prime focus, HSC has already been in science operation
with the wide field of view and the reasonably flat focal plane
provided by the new Wide-Field Corrector lens system (WFC). The WFC
will be used for PFS as well. Mechanically, the new prime focus
housing unit ``POpt2'' is integrated with WFC and accommodates the HSC
instrument inside. When PFS is in operation, the HSC instrument will
be taken out and our {\bf Prime Focus Instrument (PFI)} will be
installed in POpt2.

PFI has been developed by the collaboration of CIT\footnote{California
  Institute of Technology} \& NASA JPL\footnote{Jet Propulsion
  Laboratory}, LNA\footnote{Laborat\'{o}rio Nacional de
  Astrof\'{i}sica (Brazil)}, and ASIAA\footnote{Academia Sinica
  Institute of Astronomy and Astrophysics (Taiwan)}, accommodating key
subcomponents such as the fiber positioner system, science \& fiducial
fiber system, Acquisition \& Guide (AG) cameras, and calibration
system. The fiber positioner system consists of 42 modules each of
which accommodates 57 ``Cobra'' rotary actuators populated with
science fibers. The tip of each science fiber is equipped with a
plano-concave microlens to increase the focal ratio of the input beam
to the fiber to 2.8\cite{takato14}. The Cobra engineering model
actuators have been assembled to a prototype module and tested. The
results show satisfactory target convergence performance in the patrol
field of each fiber\cite{fisher14}. These subcomponents will be
integrated into PFI and be fully tested at ASIAA before delivery to
Subaru\cite{wang16pfi}.

{\bf Metrology Camera System (MCS)} is under development at
ASIAA\cite{wang16mcs}. It will be installed at the Cassegrain focus of
the telescope. Because the fiber positioners have no encoders, an
external system is required to drive them to the proper position.  MCS
corresponds to this external system which takes images of the science
and fiducial fibers back-lit from the other side of prime focus, and
then measures the fiber positions, enabling closed-loop operation of
the positioners. MCS is capable of taking an image of all the back-lit
science and fiducial fibers on the prime focus in one exposure. The
fiber configuration time is significantly shorter than FMOS for which
a small CCD camera needs to scan the field of view to measure all the
fibers. The 380mm aperture system is designed to minimize the impacts
of the dome seeing effect and small-scale figure errors of the WFC
lens surface shapes.

{\bf Spectrograph System (SpS)} will be integrated at
LAM\footnote{Laboratoire d'Astrophysique de Marseille}\cite{madec16},
with the fiber system delivered by LNA and the camera dewars \&
detectors developed by Princeton University (PU)\cite{gunn16} and
Johns Hopkins University (JHU)\cite{smee16, hart16}. The divergent
beams from the science fibers on the pseudo slit are collimated and
then split into blue, red and NIR channels by two dichroic
mirrors. After this, the beam is dispersed by the VPH grating and
spectral images are formed on the detectors. A grating exchange
mechanism allows a higher dispersion VPH grism to be accommodated in
the system and deliver medium resolution spectra in the red channel
with no changes in the other parts of SpS. SpS consists of four
spectrograph modules (SM) each of which is identically designed to
deliver $\sim$600 spectral images on the detectors.

{\bf Fiber system ``FOCCoS''} to be delivered by LNA\cite{cesar14}
consists of three parts: Two short-fiber systems accommodated in PFI
and SpS, and a long cable system between them routed on the
telescope. The route of this long one is still being finalized, but
the total fiber length will be approximately 65m. These three parts
are connected together by two sets of fiber connectors. One is needed
at the telescope top end to make POpt2 detachable from the telescope,
and the other is in front of SpS to ease the delivery and integration
of SpS at Subaru and to make the operation and maintenance activities
independent of the other PFS subsystems.

In Table \ref{tab:params}, the major instrument parameters are
listed. While the basic parameters are fixed, one should refer to the
PFS official web site \url{http://pfs.ipmu.jp/} for the fiber
reconfiguration time, the throughput and the estimated on-sky
sensitivity and related details as they will be updated as the
instrument is built, integrated and tested and the characteristics are
better understood.

\begin{table}[!ht]
  \caption{PFS major instrument parameters}
  \label{tab:params}
\smallskip
\begin{center}
{\small
\begin{tabular}{|c|c|c|c|c|} \hline
\multicolumn{5}{|c|}{\bf Prime Focus Instrument (PFI)} \\ \hline
\multirow{2}{*}{Field of view (hexagonal)} & \multicolumn{4}{c|}{Diameter of circumscribed circle: 1.38 deg} \\
 & \multicolumn{4}{c|}{Area: 1.25 deg$^2$} \\ \cline{1-5}
 Number of fibers & \multicolumn{4}{c|}{2394 science fibers and 96 fixed fiducial fibers.} \\ \cline{1-5}
Fiber density & \multicolumn{4}{c|}{2000 deg$^{-2}$ (0.6 arcmin$^{-2}$)} \\ \hline
Fiber core diameter & \multicolumn{4}{c|}{127$\mu$m ($=$1.12 (1.02) arcsec
     at the field center (edge), respectively)} \\ \hline
Positioner pitch & \multicolumn{4}{c|}{8mm ($=$90.4 (82.4) arcsec at the field center (edge), respectively)} \\ \hline
Positioner patrol field diameter & \multicolumn{4}{c|}{9.5mm ($=$107.4
     (97.9) arcsec at the field center (edge), respectively)} \\ \hline
Fiber minimum separation & \multicolumn{4}{c|}{$\sim$30 arcsec} \\ \hline
Fiber configuration time & \multicolumn{4}{c|}{$\sim$60-100 sec (TBC)} \\ \hline
Number of AG cameras & \multicolumn{4}{c|}{6} \\ \hline
Field of view per AG camera & \multicolumn{4}{c|}{5.1 arcmin$^2$} \\ \hline
Sensitivity of AG camera & \multicolumn{4}{c|}{$S/N=$30(100) for $r=20$ mag (AB), 1(10) sec exposure.} \\ \hline
\multicolumn{5}{|c|}{\bf Spectrograph System (SpS)} \\ \hline
\multirow{2}{*}{Spectral arms} & \multirow{2}{*}{Blue} &
         \multicolumn{2}{c|}{Red} & \multirow{2}{*}{NIR} \\ \cline{3-4}
 & & Low Res. & Mid. Res. & \\ \hline
Spectral coverage & 380-650nm & 630-970nm & 710-885nm & 940-1260nm \\ \hline
Dispersion & 0.7 \AA$/$pix & 0.9 \AA$/$pix & 0.4 \AA$/$pix & 0.8 \AA$/$pix \\ \hline   
Spectral resolution & 2.1 \AA & 2.7 \AA & 1.6 \AA & 2.4 \AA \\ \hline  
Resolving power & 2300 & 3000 & 5000 & 4300 \\ \hline  
SpS throughput & 53\% (at 500nm) & 57\% (at 800nm) & 54\% (at 800nm) &
 33\% (at 1100nm) \\ \hline
\end{tabular}
}
\end{center}
\end{table}

\section{OPERATION CONCEPT}
\label{sec:operation}

For the operation of the PFS instrument and the large survey program
it will carry out, coordination not only between the hardware and
software but also between different software packages is crucial. Four
software components are under development for this, with different
sets of functions packaged and therefore designed to be only loosely
coupled to each other. The detailed definitions of these four packages
have been evolving as the instrument and survey operation concepts are
being updated to maximally accommodate the distinct features of the
planned survey for PFS SSP such as: (1) A much fainter limit than
previous legacy surveys such as SDSS is pursued, exploiting the large
light-gathering power of the Subaru Telescope, (2) given the wide
variety of scientific goals, a wide variety of objects are targeted
for observation and therefore a variety of definitions of success need
to be encompassed, (3)
PFS allows dynamic fiber reallocation even on an individual exposure
basis, unlike the static integration in case of classical multi-slit
and multi-fiber spectroscopy using machined plates. In addition, since
PFS will be a facility instrument at the Subaru Telescope observatory
and will be operated in the framework of general open-use observation,
we are continually discussing all aspects of operations with the
observatory and are trying to adapt our plans accordingly. Below an
overview is given of the current operation concepts and the main
bodies of the software definitions. Technical details are covered in
another article\cite{shimono16}.

\subsection{Observation preparation}

The observation process starts with preparing an input target catalog
which includes not only science targets but also stars for field
acquisition, auto-guiding \& focusing, and flux calibration.  Then
telescope pointings, position angles, and fiber allocations to science
targets at each pointing are defined for a given time of
observation. This planning task is performed by a software package
called {\bf ``ETS'' (Exposure Targeting Software)} being developed by
MPA\footnote{Max-Planck-Institut f\"{u}r Astrophysik}. This
observation configuration is prepared by scientists at sites off the
observatory, and the prepared configuration is then uploaded to a {\bf
  database} system. At the actual time of observation, which may be
different from the one in the plan, the details in the mapping between
the science targets and allocated fibers are recalculated before the
fiber configuration starts. Observers should assign a fraction of
fibers to observe blank sky regions and another fraction to observe
flux calibration stars simultaneously with the science targets. The
optimal numbers and distributions of these fibers for sky and
calibration targets will be determined in the course of on-sky
engineering observations.

\subsection{Instrument operation \& data acquisition}

{\bf ``ICS'' (Instrument Control Software)} is the software package
that orchestrates the PFS subsystems and subcomponents for instrument
operation in coordination with the telescope system. The key component
for the integration and coordination is a messaging hub system
(MHS). As has been demonstrated in the SDSS operations at Apache Point
Observatory, it efficiently organizes distributed processes providing
uniform communication interfaces between subcomponents. MHS is being
used for the operation of the CHARIS instrument\cite{charis} (a high
contrast integral field spectrograph for studying disks and extrasolar
planets around stars) which is now in the commissioning phase at the
Subaru Telescope, so we will take advantages of this experience in
advance of the PFS commissioning process.

In actual observations, the first step is to point the telescope to
the target field and set the instrument rotator to a requested angle.
In parallel, the Hexapod, Atmospheric Dispersion Corrector, and
telescope primary mirror active support are adjusted by the telescope
control system for this new field. Once the telescope and rotator stop
slewing and start operating in the tracking mode, exposures of
acquisition stars are taken by the A\&G cameras, errors in the
telescope pointing and rotator angle are calculated, and feedback is
sent to the telescope control system for corrections.  This process is
iterated continuously until the errors are small enough to start
auto-guiding. A focusing operation can be executed at some point in
this course of field acquisition and auto-guiding.

While the telescope is slewing, the fiber positioner system can start
configuring the science fibers at least coarsely to the expected
positions of the science targets on the focal plane.
The MCS measures the current positions of the science fibers and
errors from the requested positions. Based on this information the
fiber positions are updated and the errors get smaller as successive
iterations are applied. Once the telescope and rotator are in the
tracking mode, the rotator operation and auto-guiding are temporarily
stopped\footnote{When observations are carried out using the prime
  focus, the prime focus instrument can be rotated, but the Cassegrain
  instrument cannot be operated because the focus is not in
  use. Accordingly, the MCS cannot be rotated synchronously with PFI
  and therefore the images of back-illuminated fibers on MCS are
  elongated if PFI is rotated and could worsen the centroiding error.}
for fine positioning of the fibers (the telescope can still be moving
in the tracking mode).
Apart from the time for telescope slewing and rotator operation, one
iteration of fiber configuration is expected to take $\sim$10$-$15sec,
including both the time of Cobra moves and exposure time of MCS. We
expect $\sim$6 iterations will be required, so one fiber
configuration will be completed in $\sim$90sec given several
iterations are needed, but more studies are on-going to fully
understand the timing budget.

Once the fiber configuration is complete, the rotator operation and
the auto-guiding operation are restarted, and the spectrograph system
then starts taking exposures as requested. The data format from each
exposure are two 4K $\times$ 2K CCDs from the blue and red cameras,
and one 4K $\times$ 4K H4RG detector from the NIR camera. At the end
of each exposure, the data are read out and passed on to the Data
Reduction Pipeline (DRP) for on-site data reduction, data quality
assessment \& assurance, and data archival.

\begin{figure}
  \begin{center}
    \begin{tabular}{ccc} 
      \includegraphics[width=7.5cm]{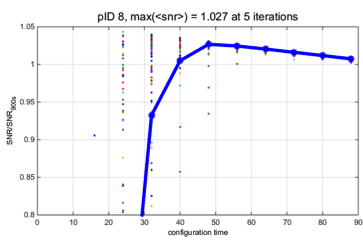} &
      \includegraphics[width=7.5cm]{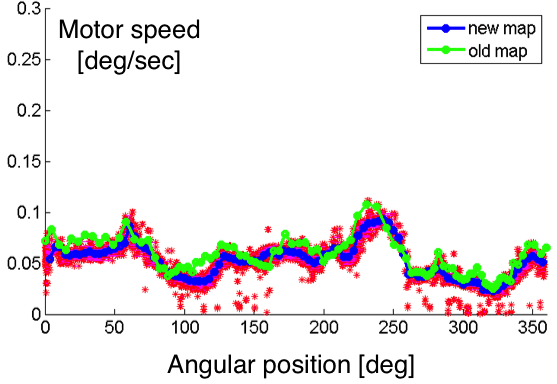} \\
    \end{tabular}
  \end{center}
  \caption{ \label{fig:cobraperformance} On the left, the level of
    convergence of Cobras to requested positions is plotted against
    measure of the signal-to-noise ratio relative to a fiducial value,
    which increases to the maximum at the fourth iterations and slowly
    decreases at later iterations. These data were taken at the target
    convergence tests on the engineering model Cobra positioners
    module. On the right, the data points indicate the Cobra motor
    speeds as a function of angular position. The relationship is
    so-called a motor map. As two curves are shown in the graph, a
    motor map can be updated as more data are collected.}
\end{figure}

Below are a few details to be highlighted in the procedure of
observation and data acquisition procedure:

\begin{itemize}
\item It is not trivial to determine that the Cobras are aligned well
  enough to the target positions. Actual criteria for convergence and
  procedure of assessment are still under discussions, but we
  consider that the concept of a signal-to-noise ratio can be useful
  as a measure of optimality, rather than the residual distances of
  the fibers from their targets. Starting from a situation in which
  all the fibers are remote from their target positions (except for
  chance coincidence), in the first few iterations, all the fibers get
  closer to the targets, so the signal in terms of fluxes from the
  objects to the fibers increases quickly. However, the gain of the
  signal after each move of the Cobras gets smaller at later
  iterations because most of the fibers are already reasonably close
  to the targets, and then the gain becomes less than the loss due to
  the loss of the observing time for integration on the
  detectors. Therefore the number of iterations giving the optimal
  signal-to-noise ratio must be somewhere in the middle
  (Fig. \ref{fig:cobraperformance}). In reality, since the time for
  one iteration is rather short in particular at later iterations,
  some more iterations to pursue better positions of more fibers may
  be worth at relatively small expenses of observing time, so we will
  leave such flexibility in the choice.
\item The response of each motor in the Cobra actuator to a drive
  signal is known to depend on angular position
  (Fig. \ref{fig:cobraperformance}). Characterizing this so-called
  ``motor map'' and operating the Cobras taking it into account are
  considered key to achieve efficient convergence.
\item Although observation strategies and data quality success
  criteria are different among the three main science areas, all three
  are planning to acquire data with no beam-switching to blank sky.
  (The instrument control system will accommodate the beam switching
  capability as an option.) This will maximize the on-source
  integration time over the observing time and minimize the geometric
  constraints in the allocation of science fibers to science
  targets\footnote{In cross-beam switching observations, two fibers
    are assigned to one science target and the telescope pointing is
    dithered between one exposure and another so that in the first
    exposure one of the fibers is placed on the target and the other
    observes blank sky, and in the next exposure, they switch the
    role. This way, 100\% of the exposure time can be used for
    on-source integration, but the fibers can be significantly less
    flexibly allocated to targets.}.
 \item Due to the large field of view, the differential atmospheric
   refraction effect is severe over most of the sky. Accordingly, the
   science fibers need to be reconfigured e.g. every $\sim$30 minutes
   (TBC).
 \item The instrument rotator operates over a restricted range between
   $-60$ deg to $+60$ deg. This is to minimize any variation of the
   fiber status by instrument rotation (possibly important for stable
   Point Spread Function (PSF) on the spectrograph detectors and
   therefore for good sky subtraction), exploiting the hexagonal
   symmetry on the PFS focal plane.
\end{itemize}

\subsection{Data reduction \& spectral calibration strategy}

{\bf ``DRP'' (Data Reduction Pipeline)} comprises the ``2D'' part
(2D-DRP) and the ``1D'' part (1D-DRP). The 2D-DRP, which is under
development by PU, receives two-dimensional raw spectral FITS images
read out from the detectors and produces one-dimensional,
sky-subtracted, flux- and wavelength-calibrated spectra ready for
scientific analyses. The 1D-DRP being developed by LAM then receives
these 1D spectra and measures various parameters of spectroscopic
features such as redshifts and emission line fluxes. After each
exposure of the spectrograph detectors, the data will be processed by
on-site DRP with calibration data sets taken in advance.
1D-DRP is applied to the reduced and calibrated 1D spectra and
measured parameters are added to the database. As successive exposures
are taken for the same objects at different nights and observation
runs, deeper and higher quality spectra will be produced from from
full, batch processing of all available data.

One important challenge for this project is to achieve the goal of sky
subtraction accuracy (down to $\sim$0.5\% in the faint sky continuum
between the lines). This means, given that we are not doing
beam-switching operations, that the sky spectrum of an object fiber
needs to be accurately modeled from the spectra of other fibers
looking at the sky, and for this, the two-dimensional fiber PSF needs
to be well characterized as functions of x and y on the detectors
(corresponding to fibers and wavelengths approximately). In other
words, the conditions during the calibration data acquisition need to
closely mimic the observing conditions at night. We have the following
plans for this:

\begin{itemize}
 \item We have a calibration lamp system on top of PFI for both
   flat-fielding and wavelength calibration. This lamp illuminates a
   quasi-Lambertian (TBC) screen on the ceiling of the telescope
   enclosure and the illumination reflects back to the telescope
   primary mirror. In this way, the telescope pupil can be diffusely
   illuminated for calibration mimicking the illumination by the sky in
   real observations at night.
 \item Even if the pupil illumination is managed as above, differences
   of the fiber status between observations and calibration may cause
   some errors in the PSF modeling due to e.g. variation of Focal
   Ratio Degradation (FRD) in the fibers. There are three cases where
   such errors may be introduced: (1) Fiber moves (mainly twists) as
   the Cobras move, (2) coils/uncoils of fiber bundles with the
   rotation of the instrument by the rotator, and (3)
   bending/unbending of the fiber cable due to telescope elevation
   changes. We currently think that (1) has the most significant
   impact, so the procedure is likely to take the calibration data for
   every fiber configuration taken in a given night. Detailed studies
   are underway. If (2) is also significant, the rotator angle should
   also be reproduced in this data acquisition but the amount of
   calibration data required could be huge and it may not be realistic
   to take them all during a night. If (2) and/or (3) are significant,
   we will plan to have another calibration lamp system to take data
   in the daytime as functions of telescope elevation and rotator
   angle and characterize the impacts.
 \item A significant number of fibers should be assigned as sky fibers
   and should be roughly uniformly distributed over the focal
   plane. The required number is still TBD, and will be clarified in
   commissioning observations.
\end{itemize}

\subsection{Data quality assessment and assurance for long-term survey processing}

The procedure of data quality assessment and assurance (QA) is still
being actively discussed in detail, but we are aiming to accommodate a
data QA on an object-by-object basis: Data quality assessment
procedure and success criteria are set for each object so that, once a
particular object is considered ``done'', the fiber(s) assigned to the
object can be allocated to a different target and be reconfigured
accordingly. Also, discussions of a data QA in a short time scale
(much shorter than one night) are underway, for which ``on-site''
(i.e. at the telescope) quick data reduction is needed in addition to
``off-site'' full reduction. In particular for faint objects, one of
the key processes is full analysis of sky fibers even in the quick
on-site data QA to understand the noise characteristics and
subsequently limiting fluxes as a function of wavelength. The database
is then updated with such information, revisions are applied to the
field definitions and/or fiber configurations accordingly, and
observations are performed using such updated fields and fiber
configurations. This routine is repeated until the survey is
considered ``done''. {\bf ``SPT'' (Survey Planning and Tracking
  software)} is the software package for the survey management
responsible for data QA.

\section{RECENT DEVELOPMENTS}
\label{sec:updates}

\subsection{The collaboration}

In December 2015, a consortium of Chinese institutes\footnote{The
  members are: Shanghai Jiao Tong University, Shanghai Astronomical
  Observatory, University of Science and Technology of China, Tsinghua
  University, Xiamen University, and National Astronomical Observatory
  of China.} including 11 senior scientists by Yipeng Jing (Shanghai)
joined the PFS collaboration as a full member.
%
%
To make the collaboration even stronger and to further improve the
chance of success of instrument development and survey science, we are
still looking for new partners. There are a few groups and institutes
as candidates with which the PFS steering committee is negotiating.

\subsection{The instrument development}

\begin{figure}
  \begin{center}
    \includegraphics[width=16cm]{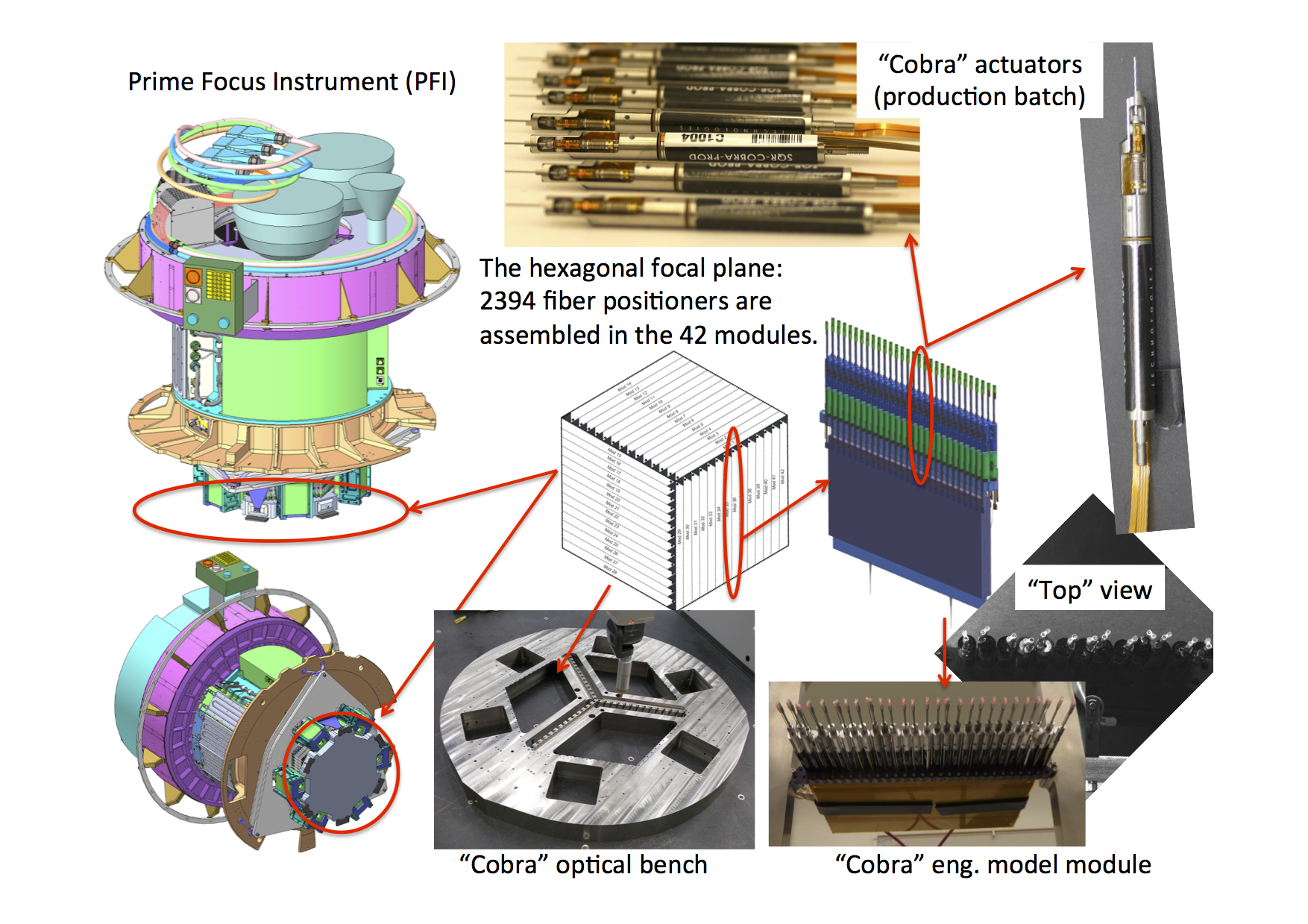}
  \end{center}
  \caption{ \label{fig:cobra} A schematic view of the PFS focal plane
    to be equipped with the 2394 fibers and Cobra actuators. Several
    photos of the real hardware components are also shown: The
    production-batch Cobra actuators (top right), a Cobra module with
    the engineering model actuators populated (bottom right), and the
    Cobra optical bench in process which is to accommodate 42 Cobra
    modules.}
\end{figure} 

\begin{figure}
  \begin{center}
    \includegraphics[width=16cm]{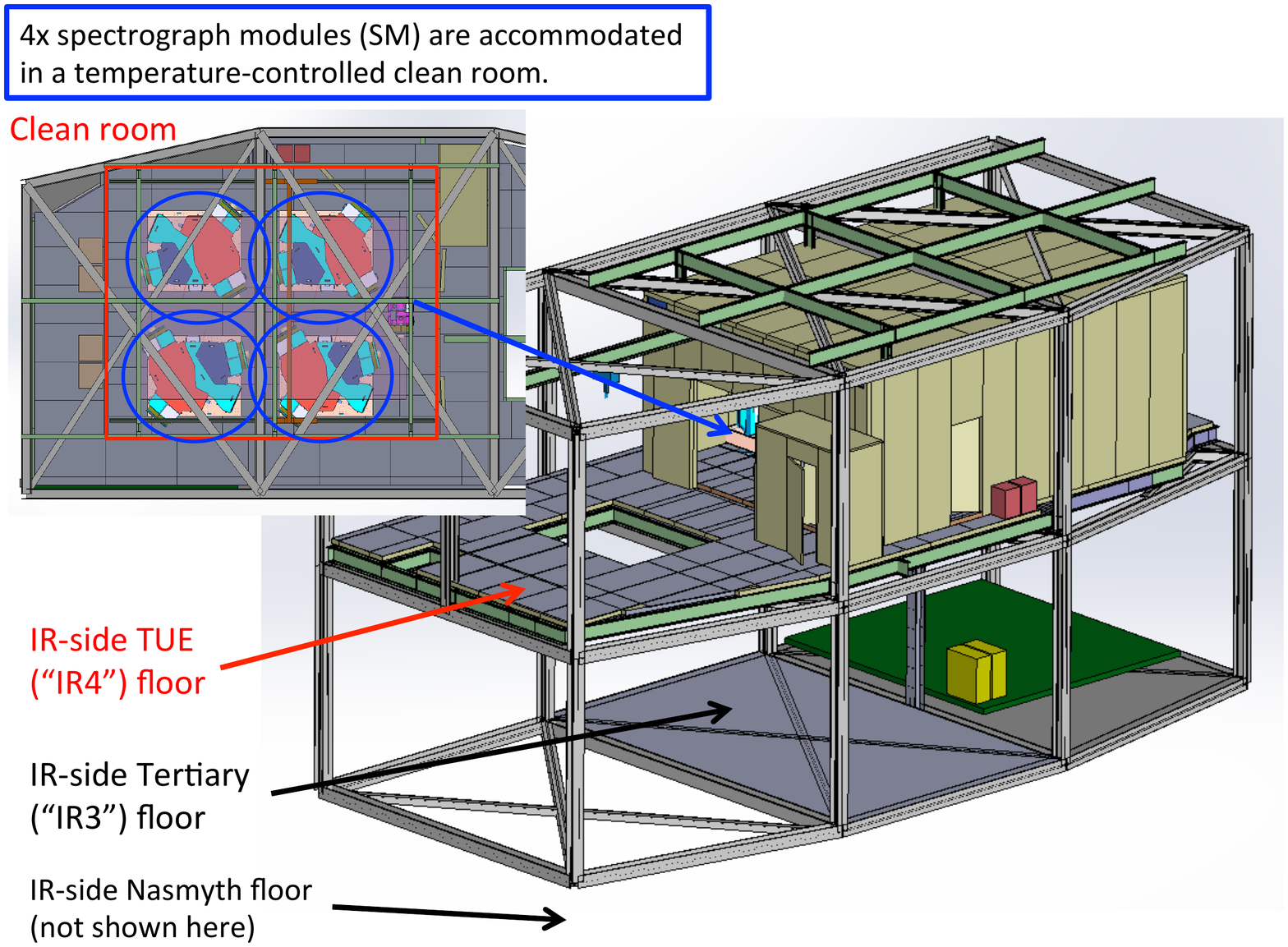}
  \end{center}
  \caption{ \label{fig:ir4scr} An overview based on the 3D model of
    the floor (IR4 floor) and temperature-controlled clean (SCR) where
    to accommodate the PFS spectrograph system.}
\end{figure} 

\begin{figure}
  \begin{center}
    \begin{tabular}{cc} 
      \includegraphics[width=7.5cm]{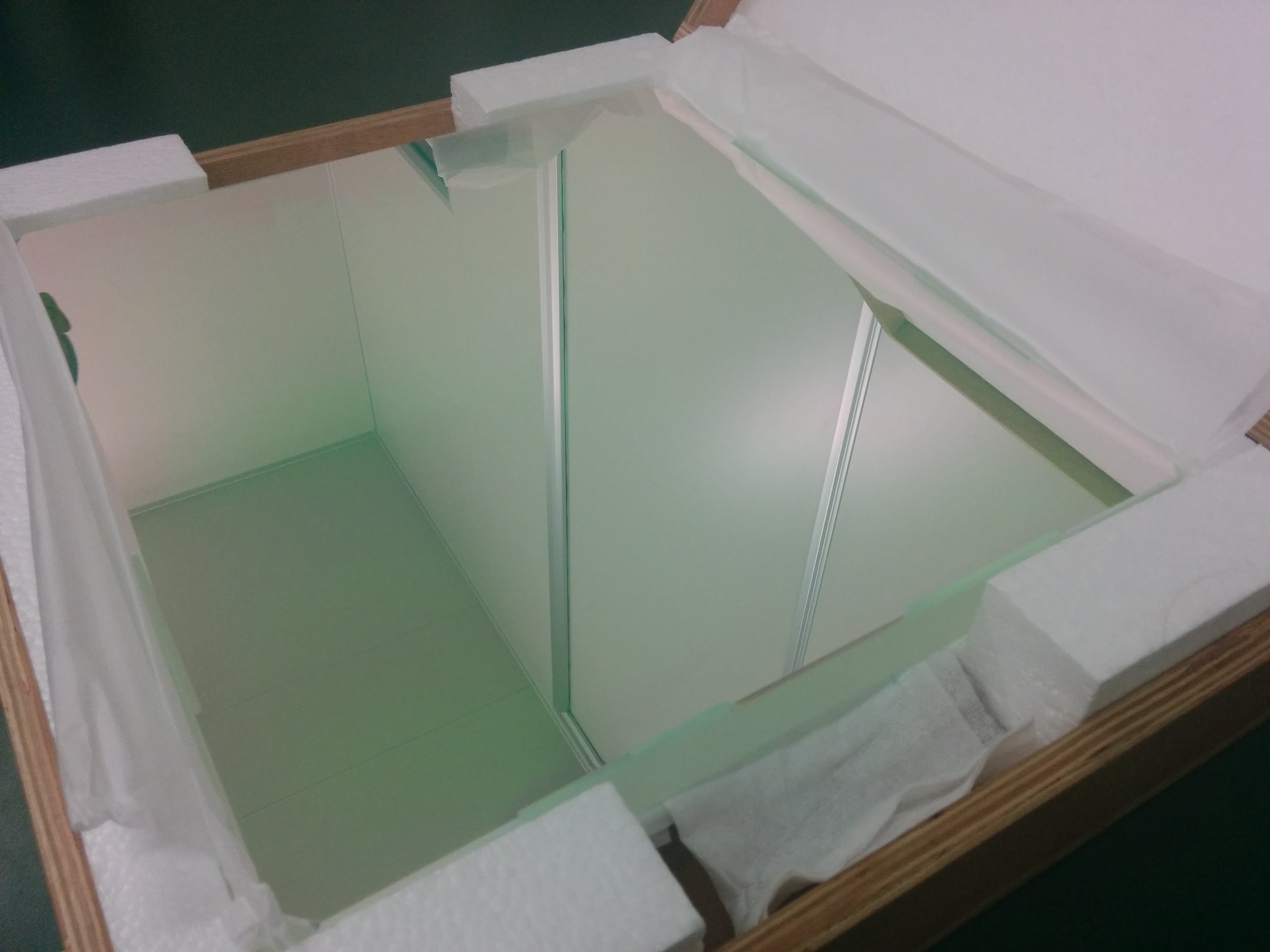} &
      \includegraphics[width=7.5cm]{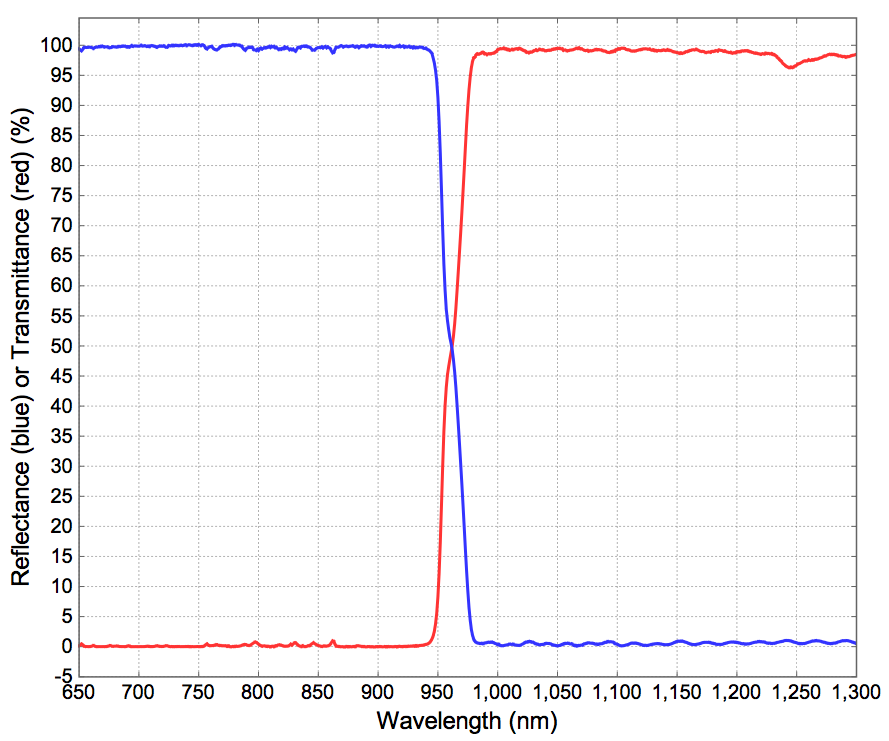} \\
    \end{tabular}
  \end{center}
  \caption{ \label{fig:red-dichroic} On the left, a photo of the first
    red dichroic mirror is shown. On the right, the measured
    reflectance (transmittance) is plotted as a function of wavelength
    by a blue (red) curve, respectively. The substrate was coated by
    Asahi Spectra Co., Ltd..}
\end{figure}

\begin{figure}
  \begin{center}
    \begin{tabular}{cccc} 
      \includegraphics[width=3.8cm]{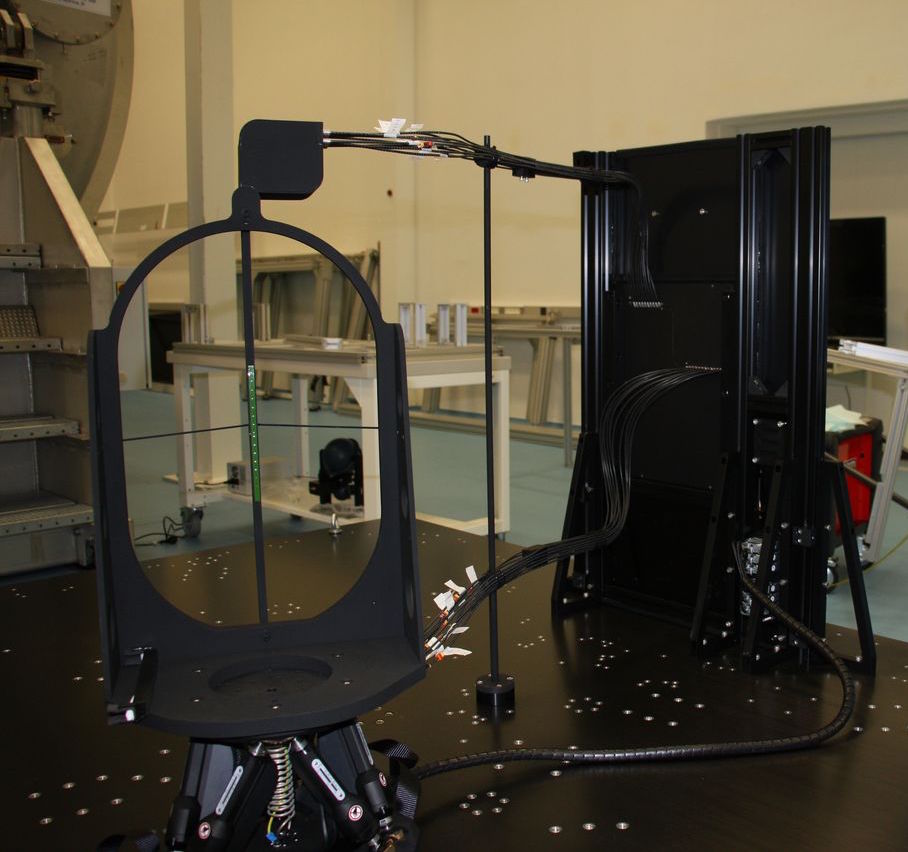} &
      \includegraphics[origin=c,angle=-90,width=3.8cm]{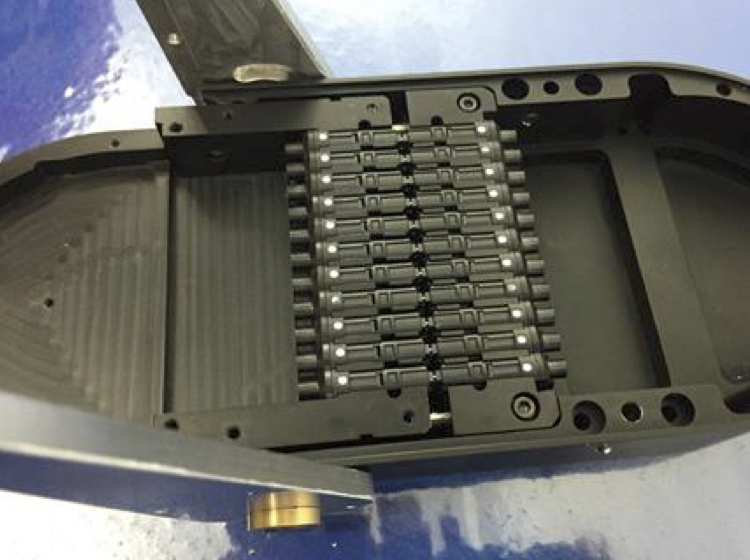} &
      \includegraphics[origin=c,angle=-90,width=3.8cm]{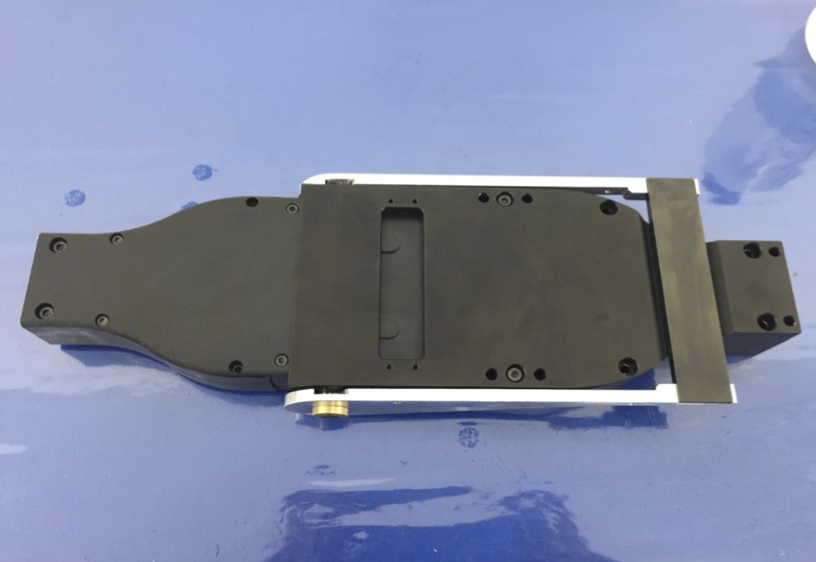} &
      \includegraphics[origin=c,angle=-90,width=3.8cm]{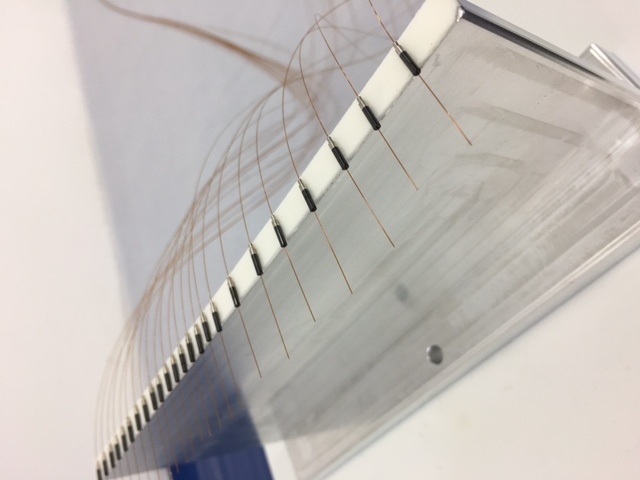} \\
    \end{tabular}
  \end{center}
  \caption{ \label{fig:lna} Photos of the parts and assemblies related
    to the fiber system\cite{cesar14}: (Left) The SpS-side fiber cable
    system, (middle-left) the inside of the large-format custom-made
    fiber connector assembly (``Tower connector'') to be installed on
    the telescope spider structure, (middle-right) the ``Tower
    connector'' when mated, and (right) the fibers and ferrules for
    the PFI-side fiber system in the integration process.}
\end{figure}

After the project passed the conceptual design review in March 2012
and the preliminary design review in February 2013, we also have gone
through critical design reviews of most of the subsystems and
subcomponents: Spectrograph system in March 2014, PFI in March 2015,
fiber positioner system in June 2015, and metrology camera in
September 2015, with additional delta reviews later when we considered
they were needed. Subsequently, the construction, integration and test
of the instrument are actively under way at the subsystem
level. Obviously, there are quite a few challenges even in the
procurement and production of individual components. Here, a few
examples are briefly introduced, while updates in the other subsystems
and subcomponents are presented in other articles by the PFS
team\cite{wang16pfi, wang16mcs, smee16, hart16, gunn16, madec16}:

\begin{itemize}
  
\item The fiber positioner ``Cobra'' (Fig. \ref{fig:cobra}): After
  several years of prototyping and testing activities\cite{fisher14},
  we recently started the mass production of the Cobras at New Scale
  Technologies. About 500 units have already been delivered as of Jun
  2016. We are extracting two units approximately every month from the
  production batch and are sending them to the life test, where a
  specific cycle of motor moves is repeated until 400k times (c.f. we
  expect about 100k cycles to be accumulated in $\sim$10 years
  operations of individual Cobras including those in engineering
  observations and calibration exposures) and the motor torques are
  measured at every 100k cycle. While one failure mode due to a rare
  manufacturing flaw of one specific parts was found in these sampled
  life tests and corrective actions were taken, the tests are
  successfully progressing and statistical evidence for durability is
  growing.

\item The detailed design of SpS platform at the Subaru Telescope
  observatory: The four SMs will be accommodated in a temperature
  controlled clean room (Spectrograph Clean Room: SCR) on the
  ``infrared-side'' 4th floor of the telescope enclosure building
  (``IR4'' floor). On this floor, the spectrograph systems of FMOS
  were operated, but the existing floor is considered inappropriate to
  accommodate PFS SpS due mainly to the significantly higher weight
  and different distribution. In 2015, design studies of the IR4 floor
  mechanical structure and SCR are intensively performed by T. Tamura
  (Subaru) (Fig. \ref{fig:ir4scr}). The FMOS spectrograph systems are
  now being dismantled and removed from the floor for the
  restructuring works to be started. Currently detailed scheduling,
  logistics and coordination after the floor is restructured and SCR
  is prepared are under discussions between the observatory and PFS
  team.

\item The dichroic coating to split the collimated beam in the
  spectrograph into the blue, red, and NIR channels: The difficulty is
  to meet the requirement of the sharp transition ($\sim$20nm) between
  the reflection and transmission as well as the high throughput in
  each regime ($\geq$95\% of reflectance and $\geq$90\% of
  transmittance), but products with satisfactory properties have
  started being delivered (Fig. \ref{fig:red-dichroic}).

\item The fiber system: Although the ``telescope part'' of the fiber
  cable system connecting PFI and SpS is still being prototyped, the
  integration and test are under way for the other parts which connect
  the PFI and SpS (Fig. \ref{fig:lna}). The procedure for implementing
  optimal quality control is being finalized.

\end{itemize}

\subsection{Instrument characterization}

\begin{figure}
  \begin{center}
    \includegraphics[width=17cm]{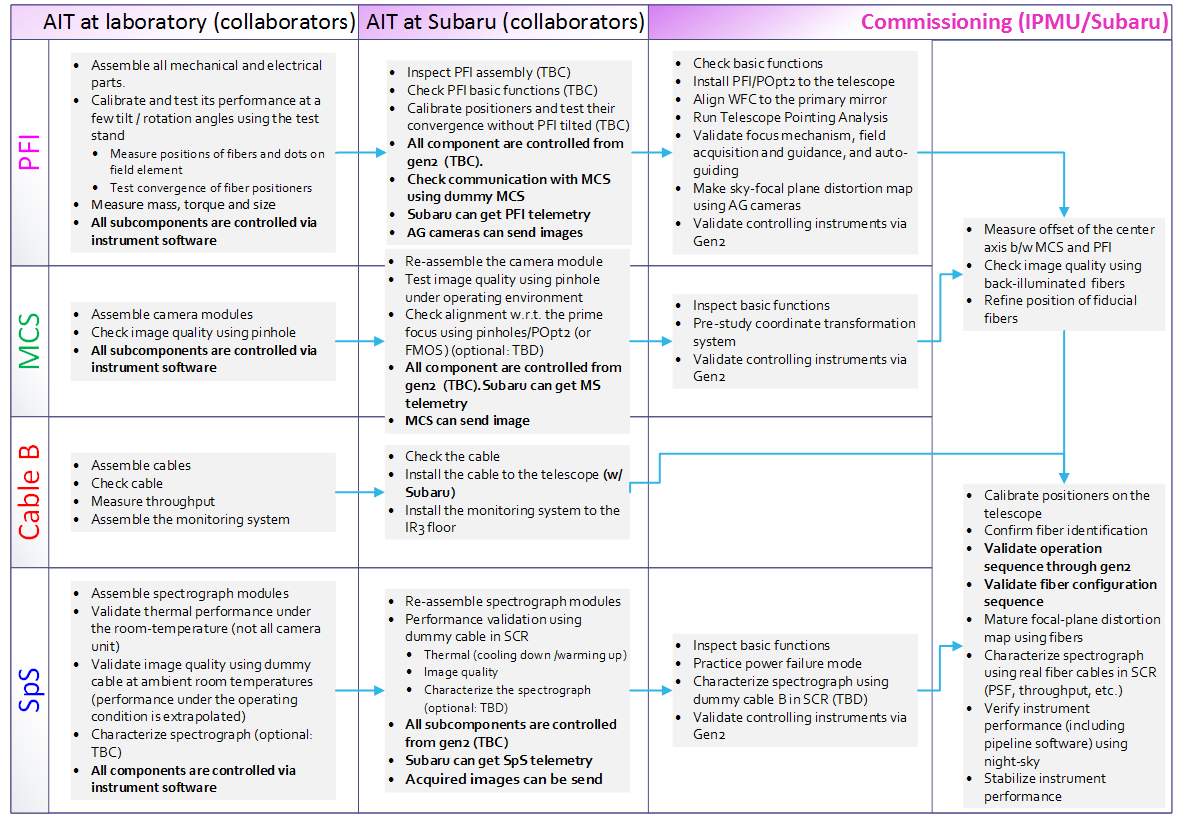}
  \end{center}
  \caption{ \label{fig:commissioning} One representation of the
    commissioning process, highlighting the transition from the
    subsystem AIT (Assembly, Integration and Test) on the left, to
    engineering observations on the right. As the commissioning
    progresses, the work moves from the collaborators to the PFS
    project office and the observatory. As the subsystems are
    integrated into the system, the tasks for system-level operation
    and performance validation increase.}
\end{figure}

In parallel to these construction, integration and test activities for
the subsystems, the detailed plans of the commissioning process are
under development (Fig. \ref{fig:commissioning}). The commissioning is
divided in two parts: the re-integration and test of the subsystems at
the observatory and engineering observations on the telescope. The
plan needs to be elaborated by discussions with the observatory for
detailed configurations with various resources and constraints at the
observatory. It should also be optimized in terms of efficiency making
sure that we will take full advantage of the complementarity with the
advanced subsystem-level integration and test processes. We are now
listing the tasks that can be completed off the telescope and those to
be done on the telescope, coordinate them along a time sequence
according to the dependencies between them, and develop a schedule of
engineering observations which will likely be separated into several
runs each of which will span a few to several nights. One of the
important works at the beginning of engineering observations is to
understand the way of operating the Hexapod of POpt2 for the alignment
of PFI with WFC, initially by using the AG cameras only. Y. Tanaka
(Subaru) defined efficient diagnostics of tilt and de-center and is
developing a routine of taking AG camera images of bright stars and
applying corrections so that the alignment can be achieved in a few
iterations. In the next couple of years, we will try to adaptively
mature the plan as the subsystem integration and test at the subsystem
level progress.

\begin{figure}
  \begin{center}
    \begin{tabular}{cc} 
      \includegraphics[width=8cm]{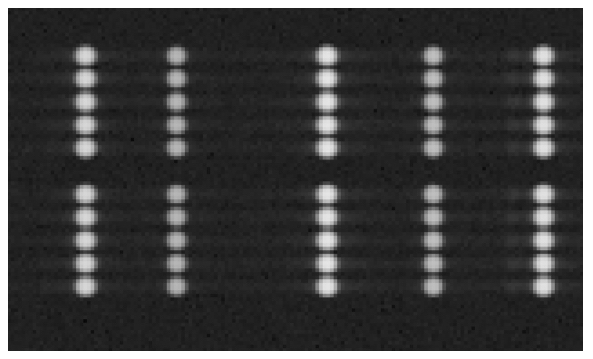} &
      \includegraphics[width=8cm]{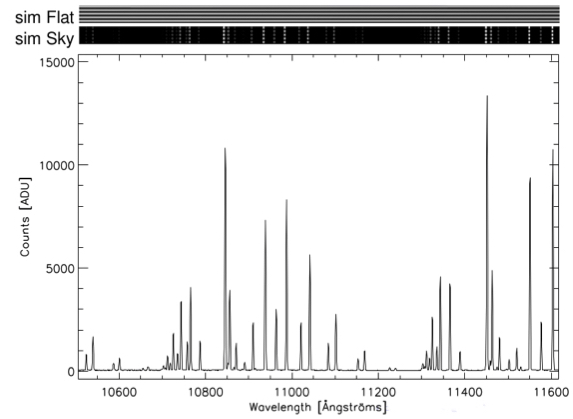} \\
    \end{tabular}
  \end{center}
  \caption{ \label{fig:simimg} Examples of simulated images and
    spectra for the 2D-DRP development: a few Neon and Argon lines on
    the left, and OH night-sky lines on the right.}
\end{figure}

For the studies of on-sky sensitivity and survey planning, a spectrum
simulator has been prepared by C. Hirata (Ohio State), K. Yabe (Kavli
IPMU), and R. Lupton (Princeton), modifying the exposure time
calculator for the WFIRST project\cite{hirata12} with the relevant set
of information for PFS including an updated throughput model. The
output from this simulator has been matched with the preliminary data
model for PFS that has recently been defined based on that of SDSS,
and it will be used to develop and validate the software components
such as 1D-DRP and the database. Meanwhile, the Princeton team has
been developing a simulator of raw images from the spectrograph
detectors where the ray-trace calculations of the optical model and
detector characteristics such as bias and dark are considered. These
simulated 2D images (Fig. \ref{fig:simimg}) and spectra have been used
to develop and validate the 2D-DRP. They have coded a prototype
pipeline, with which the data from the real spectrograph module will
be reduced and analyzed during the integration and test at LAM. The
data will be crucial not only to characterize the spectrograph module
but also to develop the algorithms of 2D-DRP for data modeling such as
PSF characterization.

\section{SUMMARY AND FUTURE PERSPECTIVES}
\label{sec:future}

PFS (Prime Focus Spectrograph), a next generation facility instrument
on the Subaru Telescope, is a very wide-field, massively multiplexed,
optical and NIR spectrograph: The prime focus will be equipped with
2394 reconfigurable fibers in the 1.3 deg field of view, and the
spectra simultaneously cover the wide range of wavelengths from 380nm
to 1260nm at one exposure. The development of this instrument by an
international collaboration under the initiative of Kavli IPMU is
finalizing the design and starting the construction at the subsystem
level. We expect the subsystems to be integrated and validated by the
collaborators and be delivered to the telescope site in 2017-2018. We
will then carry out system integration and start engineering
observations from early 2018. Based on our preliminary plans,
$\sim$1.5 years will be necessary to complete engineering observations
(including a certain period for optimization and stabilization of the
performance and operation), so we expect to start science operation
and a 5-year PFS SSP survey from mid-late 2019. The science team is
developing a detailed survey strategy to be refined in the next two
years, and the technical team is committing to this by brushing up the
estimates of on-sky instrument sensitivity and carrying out survey and
data simulations. Information on the instrument development and survey
strategy will be posted and updated on the PFS official website
\url{http://pfs.ipmu.jp/}. In addition, news, events and milestones
are reported in the PFS official blog \url{http://pfs.ipmu.jp/blog/}.

PFS and HSC, a unique set of powerful survey instruments, will be
crucial strategic pieces for the Subaru Telescope through the 2020s
into the 2030s, allowing unique science by effective synergies with
new generation ground-based and space missions such as TMT, LSST,
Euclid and WFIRST.

\acknowledgments 

We appreciate all the contributions from the PFS science team to the
instrument requirements definitions and the survey planning. We also
thank the people involved with this PFS project in the past in any
formats. Without their efforts and contributions, the project would
not even exist now. We are grateful to the staffs (in addition to
those on the author list) at National Astronomical Observatory of
Japan and the Subaru Telescope observatory for their contributions to
the development of the PFS instrument, the modifications of the
telescope system and other infrastructures to accept PFS, the
preparations of PFS system integration and engineering observations,
and various other aspects such as the administrative supports. Our
thanks should also go to the staffs at Durham University, UK, for
their supports of the development of the PFS fiber system as the
consultancy. We are grateful to the external reviewers at the PFI
critical design review (Kim Aaron, Mark Colavita, Randy Foehner, Kirk
Seaman from JPL, French Leger from University of Washington, and Ted
Huang from ASIAA) for their insightful and valuable inputs. We
gratefully acknowledge support from the Funding Program for
World-Leading Innovative R\&D on Science and Technology (FIRST)
program "Subaru Measurements of Images and Redshifts (SuMIRe)", CSTP,
Japan. This work is supported by JSPS KAKENHI Grant Numbers
JP15H05893, JP15K21733, and JP15H05892. The work in ASIAA, Taiwan, is
supported by the Academia Sinica of Taiwan. The work in Brazil is
supported by the FAPESP grant 2012/00800-4.

\bibliography{ntamura-pfs} 

\begin{thebibliography}{10}

\bibitem{miyazaki02}
Miyazaki, S., Komiyama, Y., Sekiguchi, M., Okamura, S., Doi, M., Furusawa, H.,
  Hamabe, M., Imi, K., Kimura, M., Nakata, F., Okada, N., Ouchi, M., Shimasaku,
  K., Yagi, M., and Yasuda, N., ``Subaru {P}rime {F}ocus {C}amera --
  {S}uprime-{C}am,'' {\em PASJ}~{\bf 54},  833--853 (2002).

\bibitem{kimura10}
Kimura, M., Maihara, T., Iwamuro, F., Akiyama, M., and Tamura, N.~{\it et al}.,
  ``Fibre {M}ulti-{O}bject {S}pectrograph ({FMOS}) for the {S}ubaru
  {T}elescope,'' {\em PASJ}~{\bf 62},  1135--1147 (2010).

\bibitem{tamura12}
Tamura, N., Takato, N., Iwamuro, F., Akiyama, M., and Kimura, M.~{\it et al}.,
  ``Subaru {FMOS} now and future,'' in [{\em Ground-based and Airborne
  Instrumentation for Astronomy IV}{\nolinebreak\hspace{0.1em}]},  Ian
  S.~McLean, Suzanne K.~Ramsay, H.~T., ed., {\em Proc. SPIE} {\bf 8446},
  8446M0 (2012).

\bibitem{jurek04}
Brzeski, J.~K., Gillingham, P., Correll, D., Dawson, J., Moore, A.~M., Muller,
  R., Smedley, S., and Smith, G.~A., ``Echidna: the engineering challenges,''
  in [{\em Ground-based Instrumentation for
  Astronomy}{\nolinebreak\hspace{0.1em}]},  Moorwood, A. F.~M. and Masanori,
  I., eds., {\em Proc. SPIE} {\bf 5492},  1228--1242 (2004).

\bibitem{yabe14}
Yabe, K., Ohta, K., Iwamuro, F., Akiyama, M., and Tamura, N.~{\it et al}.,
  ``The mass-metallicity relation at $z \sim 1.4$ revealed with
  {S}ubaru/{FMOS},'' {\em MNRAS}~{\bf 437(4)},  3647--3663 (2014).

\bibitem{muzic12}
Mu\v{z}i\'{c}, K., Scholz, A., Geers, V., Jayawardhana, R., and Tamura, M.,
  ``Substellar {O}bjects in {N}earby {Y}oung {C}lusters ({SONYC}). {V}. {N}ew
  {B}rown {D}warfs in $\rho$ {O}phiuchi,'' {\em ApJ}~{\bf 744(2)},  134, 11 pp.
  (2012).

\bibitem{tonegawa15}
Tonegawa, M., Totani, T., Okada, H., Akiyama, M., and Dalton, G. B.~{\it et
  al}., ``The {S}ubaru {FMOS} galaxy redshift survey ({F}ast{S}ound). {I}.
  {O}verview of the survey targeting h$\alpha$ emitters at $z \sim 1.4$,'' {\em
  PASJ}~{\bf 67(5)},  id.8112 pp. (2015).

\bibitem{okumura16}
Okumura, T., Hikage, C., Totani, T., Tonegawa, M., and Okada, H.~{\it et al}.,
  ``The {S}ubaru {FMOS} galaxy redshift survey ({F}ast{S}ound). {IV}. {N}ew
  constraint on gravity theory from redshift space distortions at $z \sim
  1.4$,'' {\em PASJ}~{\bf 68(3)},  id.3824 pp. (2016).

\bibitem{miyazaki12}
Miyazaki, S., Komiyama, Y., Nakaya, H., Kamata, Y., and Doi, Y.~{\it et al}.,
  ``Hyper {S}uprime {C}am,'' in [{\em Ground-based Instrumentation for
  Astronomy}{\nolinebreak\hspace{0.1em}]},  Moorwood, A. F.~M. and Masanori,
  I., eds., {\em Proc. SPIE} {\bf 5492},  1228--1242 (2012).

\bibitem{np-desi}
Levi, M., Bebek, C., Beers, T., Blum, R., Cahn, R., Eisenstein, D., Flaugher,
  B., Honscheid, K., Kron, R., Lahav, O., McDonald, P., Roe, N., and Schlegel,
  D., ``The {DESI} {E}xperiment, a whitepaper for {S}nowmass 2013.''
  arXiv:1308.0847.

\bibitem{np-wfirst-euclid}
Spergel, D., Gehrels, N., Baltay, C., Bennett, D., Breckinridge, J., Donahue,
  M., Dressler, A., Gaudi, B.~S., Greene, T., Guyon, O., Hirata, C., Kalirai,
  J., Kasdin, N.~J., Macintosh, B., Moos, W., Perlmutter, S., Postman, M.,
  Rauscher, B., Rhodes, J., Wang, Y., Weinberg, D., Benford, D., Hudson, M.,
  Jeong, W.-S., Mellier, Y., Traub, W., Yamada, T., Capak, P., Colbert, J.,
  Masters, D., Penny, M., Savransky, D., Stern, D., Zimmerman, N., Barry, R.,
  Bartusek, L., Carpenter, K., Cheng, E., Content, D., Dekens, F., Demers, R.,
  Grady, K., Jackson, C., Kuan, G., Kruk, J., Melton, M., Nemati, B., Parvin,
  B., Poberezhskiy, I., Peddie, C., Ruffa, J., Wallace, J.~K., Whipple, A.,
  Wollack, E., and Zhao, F., ``Wide-{F}ield {I}nfrar{R}ed {S}urvey
  {T}elescope-{A}strophysics {F}ocused {T}elescope {A}ssets {WFIRST-AFTA} 2015
  {R}eport.'' arXiv:1503.03757.

\bibitem{csfr}
Bouwens, R.~J., Illingworth, G.~D., Oesch, P.~A., Stiavelli, M., van Dokkum,
  P., Trenti, M., Magee, D., Labb^^c3^^a9, I., Franx, M., Carollo, C.~M., and
  Gonzalez, V., ``Discovery of $z \sim 8$ {G}alaxies in the {H}ubble {U}ltra
  {D}eep {F}ield from {U}ltra-{D}eep {WFC3/IR} {O}bservations,'' {\em ApJ
  Letters}~{\bf 709(2)},  L133--L137 (2010).

\bibitem{dsph1}
Walker, M.~G., Mateo, M., and Olszewski, E.~W., ``Stellar {V}elocities in the
  {C}arina, {F}ornax, {S}culptor, and {S}extans d{S}ph {G}alaxies: {D}ata
  {F}rom the {M}agellan/{MMFS} {S}urvey,'' {\em AJ}~{\bf 137(2)},  3100--3108
  (2009).

\bibitem{dsph2}
McConnachie, A.~W., ``The {O}bserved {P}roperties of {D}warf {G}alaxies in and
  around the {L}ocal {G}roup,'' {\em AJ}~{\bf 144(1)},  article id. 4, 36 pp.
  (2012).

\bibitem{takada14}
Takada, M., Ellis, R.~S., Chiba, M., Greene, J.~E., and Aihara, H.~{\it et
  al}., ``Extragalactic science, cosmology, and galactic archaeology with the
  {S}ubaru {P}rime {F}ocus {S}pectrograph,'' {\em PASJ}~{\bf 66(1)},  id.R1
  (2014).

\bibitem{takato14}
Takato, N., Tanaka, Y., Gunn, J.~E., Tamura, N., and Shimono, A.~{\it et al}.,
  ``Design and performance of a {F}\verb|/#|-conversion microlens for prime
  focus spectrograph at {S}ubaru {T}elescope,'' in [{\em Ground-based and
  Airborne Instrumentation for Astronomy V}{\nolinebreak\hspace{0.1em}]},
  Suzanne K.~Ramsay, Ian S.~McLean, H.~T., ed., {\em Proc. SPIE} {\bf 9147},
  914765 (2014).

\bibitem{fisher14}
Fisher, C., Morantz, C., Braun, D., Seiffert, M., and Aghazarian, H.~{\it et
  al}., ``Developing engineering model cobra fiber positioners for the {S}ubaru
  {T}elescope's prime focus spectrometer,'' in [{\em Advances in Optical and
  Mechanical Technologies for Telescopes and
  Instrumentation}{\nolinebreak\hspace{0.1em}]},  Ram\'{o}n~Navarro, Colin
  R.~Cunningham, A. A.~B., ed., {\em Proc. SPIE} {\bf 9151},  91511Y (2014).

\bibitem{wang16pfi}
Wang, S.-Y., Schwochert, M.~A., Huang, P.-J., Chen, H.-Y., Kimura, M., Hu,
  Y.-S., Chou, C.-Y., Chang, Y.-C., Ling, H.-H., Hsu, S.-F., Morantz, C.~N.,
  Reiley, D.~J., Mao, P.~H., Braun, D.~F., Wen, C.-Y., Yan, C.-H., Karr, J.~E.,
  Murray, G.~J., Gunn, J.~E., Tamura, N., Takato, N., and Shimono, A., ``The
  current status of prime focus instrument of {S}ubaru prime focus
  spectrograph,'' in [{\em Ground-based and Airborne Instrumentation for
  Astronomy VI}{\nolinebreak\hspace{0.1em}]},  {\em Proc. SPIE} {\bf 9908}
  (2016).

\bibitem{wang16mcs}
Wang, S.-Y., Chou, C.-Y., Huang, P.-J., Ling, H.-H., Karr, J.~E., Chang, Y.-C.,
  Hu, Y.-S., Hsu, S.-F., Chen, H.-Y., Gunn, J.~E., Reiley, D.~J., Tamura, N.,
  Takato, N., and Shimono, A., ``Metrology camera system of prime focus
  spectrograph for {S}ubaru telescope,'' in [{\em Ground-based and Airborne
  Instrumentation for Astronomy VI}{\nolinebreak\hspace{0.1em}]},  {\em Proc.
  SPIE} {\bf 9908} (2016).

\bibitem{madec16}
Madec, F., Le~Mignant, D., Barrette, R., Belhadi, M., Blanchard, P., Dohlen,
  K., Ferrand, D., Jaquet, M., Le~Fur, A., Le~Merrer, J., Pascal, S., Gunn,
  J.~E., Smee, S.~A., Tamura, N., and Shimono, A., ``{SUBARU} prime focus
  spectrograph: integration, testing and performance for the first
  spectrograph,'' in [{\em Ground-based and Airborne Instrumentation for
  Astronomy VI}{\nolinebreak\hspace{0.1em}]},  {\em Proc. SPIE} {\bf 9908}
  (2016).

\bibitem{gunn16}
Gunn, J.~E., Fitzgerald, R., Hart, M., Hope, S.~C., Loomis, C., Peacock, G.~O.,
  Golebiowski, M., Carr, M., Smee, S.~A., Tamura, N., Shimono, A., and Takato,
  N., ``Detector and control system design and performance for the {S}u{MIR}e
  {P}rime {F}ocus {S}pectrograph ({PFS}) cameras,'' in [{\em Ground-based and
  Airborne Instrumentation for Astronomy VI}{\nolinebreak\hspace{0.1em}]},
  {\em Proc. SPIE} {\bf 9908} (2016).

\bibitem{smee16}
Smee, S.~A., Gunn, J.~E., Golebiowski, M., Hope, S.~C., Madec, F., Gabriel,
  J.-F., Loomis, C., Le~fur, A., Dohlen, K., Le~Mignant, D., Barkhouser, R.,
  Carr, M.~A., Hart, M., Tamura, N., Shimono, A., and Takato, N., ``Visible
  camera cryostat design and performance for the {S}u{MIR}e {P}rime {F}ocus
  {S}pectrograph ({PFS}),'' in [{\em Ground-based and Airborne Instrumentation
  for Astronomy VI}{\nolinebreak\hspace{0.1em}]},  {\em Proc. SPIE} {\bf
  9908-333} (2016).

\bibitem{hart16}
Hart, M., Barkhouser, R.~H., Smee, S.~A., Gunn, J.~E., and Loomis, C.~P.,
  ``Detector characterization for the {S}ubaru prime focus spectrograph
  ({PFS}),'' in [{\em Ground-based and Airborne Instrumentation for Astronomy
  VI}{\nolinebreak\hspace{0.1em}]},  {\em Proc. SPIE} {\bf 9915} (2016).

\bibitem{cesar14}
de~Oliveira, A.~C., de~Oliveira, L.~S., de~Arruda, M.~V., Souza~Marrara, L.,
  and dos Santos, L. H.~{\it et al}., ``Fiber optical cable and connector
  system ({FOCC}o{S}) for {PFS}/{Subaru},'' in [{\em Advances in Optical and
  Mechanical Technologies for Telescopes and
  Instrumentation}{\nolinebreak\hspace{0.1em}]},  Ram\'{o}n~Navarro, Colin
  R.~Cunningham, A. A.~B., ed., {\em Proc. SPIE} {\bf 9151},  91514G (2014).

\bibitem{shimono16}
Shimono, A., Tamura, N., Takato, N., Yasuda, N., Suzuki, N., Loomis, C.
  nad~Lupton, R.~H., Moritani, Y., and Yabe, K., ``The survey operation
  software system development for {P}rime {F}ocus {S}pectrograph ({PFS}) on
  {S}ubaru {T}elescope,'' in [{\em Software and Cyberinfrastructure for
  Astronomy IV}{\nolinebreak\hspace{0.1em}]},  {\em Proc. SPIE} {\bf 9913}
  (2016).

\bibitem{charis}
Groff, T.~D., Kasdin, N.~J., Galvin, M., Knapp, G.~R., Carr, M.~A., Brandt,
  T.~D., Peters-Limbach, M.~A., Loomis, C.~P., Jarosik, N., Guyon, O.,
  Jovanovic, N., McElwain, M.~W., Takato, N., and Hayashi, M., ``Laboratory
  testing and commissioning of the charis integral field spectrograph,'' in
  [{\em Ground-based and Airborne Instrumentation for Astronomy
  VI}{\nolinebreak\hspace{0.1em}]},  {\em Proc. SPIE} {\bf 9908-23} (2016).

\bibitem{hirata12}
Hirata, C.~M., Gehrels, N., Kneib, J.-P., Kruk, J., Rhodes, J., Wang, Y., and
  Zoubian, J., ``The {WFIRST} {G}alaxy {S}urvey {E}xposure {T}ime
  {C}alculator.'' arXiv:1204.5151.

\end{thebibliography}
\bibliographystyle{spiebib} 

\end{document}